\documentclass[reprint,prl,nofootinbib,preprintnumbers]{revtex4-1}
\usepackage{graphicx,amsmath,color}
\newcommand{\col}{\textcolor{blue}}
\def\lsim{\mathrel{\vcenter{\hbox{$<$}\nointerlineskip\hbox{$\sim$}}}}

\begin{document}
\preprint{MAN/HEP/2012/018}
\title{Implications of purely classical gravity for inflationary tensor modes }
\author{Amjad Ashoorioon$^1$, P. S. Bhupal Dev$^2$, and Anupam Mazumdar$^{1,3}$\\}
\affiliation{$^1$~Consortium for Fundamental Physics, Physics Department, Lancaster University, LA1 4YB, United Kingdom. \\
$^2$~Consortium for Fundamental Physics, School of Physics and Astronomy,
University of Manchester, Manchester, M13 9PL, United Kingdom.\\
$^3$~Niels Bohr Institute, Copenhagen University, Blegdamsvej-17, Denmark.}

\begin{abstract}
We discuss the implications of a purely classical (instead of quantum) theory of gravity for the primordial gravitational wave spectrum generated during inflation. We argue that for a scalar field driven inflation in a classical gravity the amplitude of the gravitational wave will be too small, irrespective of its primordial seed, to be detected in any forthcoming experiments. Therefore, a positive detection of the $B$-mode polarizations in the Cosmic Microwave Background spectrum will naturally {\it confirm}  the quantum nature of gravity 
itself. Furthermore there will be no upper limit on the scale of inflation in the case of classical gravity, 
and a high-scale model of inflation can easily bypass the observational constraints. 

\end{abstract}

\maketitle

Primordial inflation is one of the most successful paradigms for the early Universe cosmology (for a review, see e.g.,~\cite{Infl-rev}), which has many observational consequences~\cite{WMAP}.  One of the predictions for inflation is the generation of stochastic primordial gravitational waves along with the matter perturbations~\cite{Starobinsky:1979ty}. Typically, matter perturbations are created from the initial vacuum fluctuations which are stretched outside the Hubble patch during inflation~\cite{Mukhanov:1981xt} (for reviews, see
\cite{Brandenberger,Lidsey:1995np}).

As in any quantum field theory in a time-dependent background, the initial choice of vacuum is typically obtained by imposing the quantum commutation relationships for creation and annihilation operators  which satisfy the Wronskian condition, while confirming that the initial quantum state is the {\it least} excited state analogous to the plane wave solution emanating from deep inside the Hubble patch~\cite{Brandenberger,Lidsey:1995np}. Similar quantum calculations exist for the gravitational waves generated during inflation in which case one directly quantizes the tensor perturbations of the metric. Since the observed temperature anisotropy in the Cosmic Microwave Background (CMB) radiation is very small: $\delta T/T \sim 10^{-5}$~\cite{WMAP}, the treatment of linearized perturbation is a very good approximation for both matter and gravity sectors.

However, the assumption that gravity should also be quantized along with the matter perturbations is not yet based
upon any  observed  phenomena~\footnote{Whether gravity is {\it truly} quantum or classical is still an open issue (see~\cite{Birrell:1982ix,carlip} for some
discussions). For a consistent treatment, we often argue that gravity must be quantized along with the matter sector. However, even if gravity is treated classically, there is a hint that it might be possible to address  non-singular blackhole and cosmological solutions, found very recently in the context of higher-order {\it infinite-derivative} theories of gravity~\cite{Biswas}. }. In fact, the entire framework of matter perturbations created during inflation can be carried out without quantizing the metric fluctuations. What it means is that a pure de Sitter background without matter cannot seed the temperature anisotropy in the CMB radiation. This can also be seen by
noticing the fact that one can consistently use a {\it choice of  gauge} where scalar perturbations of the metric can be set to zero and all the temperature anisotropy would simply arise from the matter fluctuations~\cite{Brandenberger,Lidsey:1995np}.

Therefore, a natural question arises -- what if we had to treat the primordial gravity waves purely at the classical level by assuming that the space-time is indeed classical and we just quantize the matter sector in a given space-time background.
The aim of this letter is to explore such a possibility and to show what are the differences in predictions one would expect if gravity were to be treated classically, and in particular what would  be the amplitude of the primordial gravitational waves.

We recall here that in a scalar-driven inflationary model, the tensor modes are generated from tensor fluctuations of the metric \cite{Brandenberger,Lidsey:1995np}:
\begin{equation}
ds_{T}^{2}= a^2(\tau )\left(d\tau ^{2}-[\delta_{ij}+h_{ij}]dx^{i}dx^{j}\right) \label{metric-tensor}
\end{equation}
with $|h_{ij}|\ll 1$, where $h_{ij}$ is a symmetric three-tensor field satisfying $h_{i}^{i}=0=h_{ij}{}^{,j}$ whose dynamics could be determined by expanding the Einstein-Hilbert action to second order:
\begin{equation}
S_{T}^{(1)}=\frac{M_p^{2}}{64\pi }\int d\tau d^{3}\mathbf{x}~a^{2}(\tau
)~\partial _{\mu }h^{i}{}_{j}~\partial ^{\mu }h_{i}{}^{j},  \label{tensor-action-1}
\end{equation}
where $M_p$ is the usual Planck mass.
It is possible to reformulate the tensor action given by Eq.~(\ref{tensor-action-1}) to give it the appearance of a Minkowski-space theory with variable
mass term by introducing the re-scaled variable $P^{i}{}_{j}$:
\begin{equation}
P^{i}{}_{j}(x)=\sqrt{\frac{M_p^{2}}{32\pi}}~a(\tau )h^{i}{}_{j}(x),
\label{43}
\end{equation}
whose dynamics is governed by
\begin{eqnarray}
S_{T}^{(2)}&=&\frac{1}{2}\int d\tau d^{3}\mathbf{x}\nonumber\\
&&\times \left(\partial _{\tau
}P_{i}{}^{j}\partial ^{\tau }P^{i}{}_{j}-\delta ^{rs}{\partial }%
_{r}P_{i}{}^{j}{\partial }_{s}P^{i}{}_{j}
+\frac{a^{\prime \prime }}{a}%
P_{i}{}^{j}P^{i}{}_{j}\right) \nonumber
\end{eqnarray}
and is different from $S_{T}^{(1)}$ by a total time derivative. One can decompose $P^{i}{}_{j}$ into its Fourier components:
\begin{equation}
P^{i}{}_{j}=\sum\limits_{\lambda =+,\times }\int \frac{d^{3}\mathbf{k}}{%
(2\pi )^{3/2}}~p_{\mathbf{k},\lambda }(\tau )~\epsilon ^{i}{}_{j}(\mathbf{k}%
;\lambda )~e^{i\mathbf{k}\cdot \mathbf{y}},  \label{P-Fourier-Decomposition}
\end{equation}
where the sum is over two independent polarization states, usually denoted as $\lambda =+,\times $. $\epsilon ^{i}{}_{j}(\mathbf{k};\lambda )$ is the polarization
tensor satisfying the following conditions:
\begin{eqnarray}
&&\epsilon _{ij} =\epsilon _{ji},~~
\epsilon ^{i}{}_{i} =0, ~~ k^{i}\epsilon _{ij} =0,~~ {\rm and}\nonumber\\
&& \epsilon ^{i}{}_{j}(\mathbf{k};\lambda )\epsilon ^{j\ast }{}_{i}(\mathbf{k}%
;\lambda ^{\prime }) =\delta _{\lambda \lambda ^{\prime }}.
\end{eqnarray}
It is often convenient to choose $\epsilon _{ij}(-\mathbf{k}%
;\lambda )=\epsilon _{ij}^{\ast }(\mathbf{k};\lambda )$ which implies that
\begin{equation}
p_{\mathbf{k},\lambda }=p_{-\mathbf{k},\lambda }^{\ast } \label{p-pstar}
\end{equation}
in Eq.~(\ref{P-Fourier-Decomposition}).
This brings the Einstein-Hilbert action for tensor modes to the following form:
\begin{eqnarray}
S_{T}^{(2)}&=&\sum\limits_{\lambda =+,\times }\int d\tau d^{3}\mathbf{k}\nonumber\\
&&\times \left(\left( \partial _{\tau }\left| p_{\mathbf{k},\lambda }\right|
\right)
^{2}-\left( k^{2}-\frac{a^{\prime \prime }}{a}\right) \left| p_{\mathbf{k}%
,\lambda }\right| ^{2}\right).  \label{tensor-action-Fourier-modes}
\end{eqnarray}
At this point, one can assume that the tensor perturbations during inflation are either classical or quantum-mechanical.

Let us briefly discuss what happens when gravity is treated quantum-mechanically. In this case,  the field $p_{\mathbf{k},\lambda }$ is now promoted to an operator, which can be expanded in terms of creation and annihilation operators:
\begin{equation}
\hat{p}_{\mathbf{k},\lambda }=p_{k}(\tau
)\hat{a}_{\mathbf{k},\lambda }+p_{k}^{\ast }(\tau
)\hat{a}_{\mathbf{k},\lambda }^{\dagger }\,.\label{p-a-a-dagger}
\end{equation}
The mode function $p_{k}(\tau )$ satisfies the following equation of motion:
\begin{equation}
p_{k}^{\prime \prime }+\left( k^{2}-\frac{a^{\prime \prime }}{a}\right)
p_{k}=0\,.  \label{mode-function-eom}
\end{equation}
The Fourier-transformed field $\hat{p}(\tau,\mathbf x)$ and its conjugate momentum $\hat{\pi}(\tau,\mathbf x)$ satisfy the canonical commutation relations on
hypersurfaces of constant $\tau$:
\begin{eqnarray}
&&\lbrack \hat{p}(\tau ,\mathbf{x}),\hat{p}(\tau ,\mathbf{x^{\prime }})]=0,~~[\hat{%
\pi}(\tau ,\mathbf{x}),\hat{\pi}(\tau ,\mathbf{x^{\prime }})]=0 \,, \nonumber\\
&&\lbrack \hat{p}(\tau ,\mathbf{x}),\hat{\pi}(\tau ,\mathbf{x^{\prime }}%
)]=i\delta ^{3}(\mathbf{x}-\mathbf{x^{\prime }})\,,  \label{field-pi}
\end{eqnarray}
which is equivalent to imposing the following commutation relations on the creation and annihilation operators in Eq.~(\ref{p-a-a-dagger}):
\begin{equation}
\lbrack \hat{a}_{\mathbf{k}},\hat{a}_{\mathbf{k^{\prime }}}]=[\hat{a}_{%
\mathbf{k}}^{\dag },\hat{a}_{\mathbf{k^{\prime }}}^{\dag }]=0,\quad
\lbrack
\hat{a}_{\mathbf{k}},\hat{a}_{\mathbf{k^{\prime }}}^{\dag }]=i\delta
^{3}(\mathbf{k}-\mathbf{k^{\prime }}).\label{a-hat-a}
\end{equation}
These relations enforce the following Wronskian condition on the mode function $p_k(\tau)$:
\begin{equation}
p_{k}^{\ast }(\tau)\frac{dp_{k}(\tau)}{d\tau }-p_{k}(\tau)\frac{dp_{k}^{\ast }(\tau)}{d\tau }=-i.\label{Wronskian}
\end{equation}
In a de Sitter background, where $a(\tau)=-1/(H\tau)$, $H$ being the Hubble rate of expansion of the Universe, the solution to Eq.~\eqref{mode-function-eom} is
given by
\begin{equation}
\label{u-sol}
p_k(\tau)=\alpha_k~(-\tau)^{1/2}  H_{3/2}^{(1)}(-k\tau)-\beta_{k}~(-\tau)^{1/2} H_{3/2}^{(2)}(-k\tau),
\end{equation}
where $H_{3/2}^{(1)}(x)$ and $H_{3/2}^{(2)}$ are the Hankel functions of order 3/2. In the infinite past, $k\tau\rightarrow -\infty$, the mode function $p_k(\tau)$ behaves like
\begin{equation}
p_k(\tau)=-\alpha_k \sqrt{\frac{2}{k\pi}} e^{-i k \tau}+\beta_k \sqrt{\frac{2}{k\pi}} e^{i k\tau}.
\label{simple}
\end{equation}

Using the general solution given by Eq.~(\ref{u-sol}), the Wronskian condition,  Eq.~\eqref{Wronskian}, implies that
\begin{equation}
\left|\alpha_k\right|^2 - \left|\beta_k\right|^2=\frac{\pi}{4}.\label{beta-alpha-relation}
\end{equation}
In the standard theory of cosmological perturbations, it is usually assumed that the modes approached the Bunch-Davies vacuum at infinite past (see,  e.g.,~\cite{Birrell:1982ix})~\footnote{This is based on the {\it sole} assumption that no new physics appears at very small scales (see, for example,~\cite{trans-planckian}).}
\begin{equation}
p_{k}(\tau )\rightarrow \frac{1}{\sqrt{2k}}~e^{-ik\tau }
\mbox{~~~~~~for~~} k\tau \rightarrow -\infty \label{Bunch-Davies-vac}
\end{equation}
when the wavelength of the mode is much smaller than the Hubble radius. This will result in the following values for the coefficients in Eq.~(\ref{simple}):
\begin{equation}
\alpha_k=-\frac{\sqrt{\pi}}{2}, \qquad \beta_k=0\,.
\label{abq}
\end{equation}

The power spectrum of the gravitational waves for the tensor modes can be computed in the limit that the mode is well outside the Hubble patch:
\begin{equation}
{\cal P}_T = 2 \left(\frac {32 \pi}{M_{p}^2}\right) \frac{k^3}{2\pi^2}\left|\frac{p_k(\tau)}{a(\tau)}\right|^{2}_{\frac{k}{a H}\rightarrow 0},\label{power-spectrum}
\end{equation}
where the factor of two is for the two helicities of the tensor mode. In the standard case with the amplitudes given by Eq.~(\ref{abq}) and for a de Sitter background with $k/aH=-k\tau$, we obtain from Eq.~(\ref{power-spectrum}) the following power spectrum:
\begin{equation}
{\cal P}^{\rm quantum}_T=\frac{16H^2}{\pi M_p^2},
\label{p-quantum}
\end{equation}
where $H$ denotes the Hubble expansion rate of the Universe during inflation.

Let us now pause here and ask what would be different if we had treated the gravitational waves {\em classically}.

First of all, we cannot expand a classical field $p_{\mathbf{k},\lambda }$ in terms of creation and annihilation operators as in
Eq.~(\ref{p-a-a-dagger}). However, it still satisfies the equation of motion determined by Eq.~(\ref{mode-function-eom}):
\begin{eqnarray}
p''_{\mathbf{k},\lambda}+ \left(k^2-\frac{a''}{a}\right)p_{\mathbf{k},\lambda} = 0,\label{tensor-classical-eq}
\end{eqnarray}
which in a de Sitter background has a solution similar to Eq.~(\ref{u-sol}):
\begin{eqnarray}
p_{\mathbf k,\lambda}(\tau) &=& \alpha_{\mathbf k,\lambda}(-\tau)^{1/2}H^{(1)}_{3/2}(-k\tau)\nonumber\\
&& -\beta_{\mathbf k,\lambda}(-\tau)^{1/2}H^{(2)}_{3/2}(-k\tau).
\label{cmode}
\end{eqnarray}
The main difference as compared to the quantum case is that we cannot
impose the commutation relations, Eqs.~(\ref{field-pi} and (\ref{a-hat-a}), on classical fields. As a consequence, the Wronskian condition given by Eq.~(\ref{Wronskian}) is no longer valid for classical gravitational waves. In other words, the classical amplitudes $\alpha_{\mathbf k,\lambda}$ and $\beta_{\mathbf k,\lambda}$ in Eq.~(\ref{cmode}) do not have to obey the relationship given by Eq.~(\ref{beta-alpha-relation}).

Nonetheless, one has to ensure that the mode function $P^{i}{}_{j}$ is real, i.e., $p_{\mathbf{k},\lambda }$ satisfies Eq.~(\ref{p-pstar}), which imposes the following condition on the classical amplitudes:
\begin{equation}\label{alpha-beta}
\alpha_{\mathbf k,\lambda}=-\beta^*_{-\mathbf k,\lambda}\,.
\end{equation}

In a homogeneous and isotropic background, the tensor perturbation of the metric cannot be sourced by the matter perturbations at the first order. Also, it is well known that there is no source term for the space-time metric $h_{\mu\nu}$  at linear order. It cannot be sourced even at quadratic order without a violation of the adiabaticity condition for the scalar modes~\cite{kofman2}. Therefore, in a classical gravity, there is no reason  {\it a priori} why $\alpha_{\mathbf k,\lambda},\beta_{\mathbf k,\lambda} \neq 0$, unlike in the quantum case, where the Wronskian condition given by Eq.~(\ref{beta-alpha-relation}) prevents both the Bogoliubov coefficients to be zero simultaneously. Hence in a classical gravity, the amplitude of primordial gravitational waves generated during inflation can in principle be absolutely
zero, unless the tensor perturbations are sourced somehow. Such a source for $h_{\mu\nu}$ can arise, for example, from higher order perturbations. The gravitational wave solution could then obtain a nonzero solution as these higher order corrections modify the left-hand-side of Eq.~(\ref{tensor-classical-eq}) to a
nonzero value. However the amplitude of such gravity wave spectrum is known to be negligible~\cite{Ananda:2006af} at cosmological scales~\footnote{It is also possible to create a source for the equation of motion of the tensor modes at first order perturbation theory if inflation is driven by some non-Abelian gauge field~\cite{Shahin}, but  the amplitude of such gravity wave spectrum is  slow-roll suppressed. In addition, such a model, as it stands right now, is unrealistic from the point of view of creating the relevant Standard Model degrees of freedom (see e.g.,~\cite{Infl-rev}).}.

The classical power spectrum for the gravitational waves can now be computed from the formula given by Eq.~(\ref{power-spectrum}), with $p_k(\tau)$ replaced by $p_{\mathbf k,\lambda}$ given by Eq.~(\ref{cmode}) and we obtain
\begin{equation}
{\cal P}_T^{\rm classical}=\frac{64 \left|\alpha_{\mathbf k,\lambda}+\beta_{\mathbf k,\lambda}\right|^2 H^2}{M_{p}^2 \pi^2},
\label{p-classical}
\end{equation}
where $\alpha_{\mathbf k,\lambda},~\beta_{\mathbf k,\lambda}$ can be arbitrary, as long as they satisfy the reality condition given by Eq.~(\ref{alpha-beta}).
%
%

In what follows, we discuss some of the implications of the above results for detecting tensor modes. It is generically assumed that a positive detection of primordial gravitational waves via tensor modes would naturally put a bound on the scale of inflation. It must be emphasized that this is a correct statement {\em only if} gravity were treated quantum-mechanically so that the power spectrum is solely determined by the Hubble expansion rate of the Universe during inflation as in Eq.~(\ref{p-quantum}). Requiring that it must satisfy the
current observational constraint from WMAP~\cite{WMAP}, i.e.,
\begin{eqnarray}
\label{quantum-case}
{\cal P}^{\rm quantum}_T=\frac{16H^2}{\pi M_p^2} \approx  \frac{16\rho_{\rm inf}}{M_p^4}\lsim 10^{-10},
\end{eqnarray}
we obtain an upper bound on the scale of inflation,
\begin{equation}
\rho_{\rm inf}^{1/4}\sim V_{\rm inf}^{1/4}\lsim 10^{16.6}~{\rm GeV}\,,
\end{equation}
which is coincidentally very close to the Grand Unified Theory (GUT) scale.

On the other hand, for the case of classical gravity, as shown in Eq.~(\ref{p-classical}), the gravitational power spectrum depends on 
both $\alpha_{\mathbf k,\lambda},~\beta_{\mathbf k,\lambda}$. If there is no {\em initial} condition for the classical gravitational waves,  the amplitude of the gravity-wave spectrum would be {\it absolutely} zero, {\em irrespective} of the scale of inflation. This will be the case for the most common scenarios of a scalar field driven models of inflation~\cite{Infl-rev}.

Note however that due to some reason if there exists nonzero amplitude of {\it classical } gravitational waves, i.e.
$\alpha_{\mathbf k,\lambda},~\beta_{\mathbf k,\lambda}\neq 0$, the latter can be constrained from the inflaton's energy density.
Classically the energy density of the gravitational wave, when the modes are inside the Hubble patch, i.e., for $k^2\gg a''/a$, is given by
 \begin{eqnarray}
  \langle T_{00} \rangle&=&\sum\limits_{\lambda =+,\times }\left\langle \int d\tau d^{3}\mathbf{k}
~\left(\left( \partial _{\tau }\left| p_{\mathbf{k},\lambda }\right|\right)
^{2}+ k^{2} \left| p_{\mathbf{k},\lambda }\right| ^{2}\right)\right\rangle
\nonumber\\
&\propto& M_{p}^2 \sum\limits_{\lambda =+,\times }\int_{k_i}^{k_f} k d k~\left(\left|\alpha_{\mathbf k,\lambda}\right|^2+\left|\beta_{\mathbf k,\lambda}\right|^2\right),\label{GW-energy}
\end{eqnarray}
 where the second line is obtained from the fact that $\int d\tau e^{\pm 2 i k\tau} $ over time is zero. This energy density must be negligible in comparison with the energy density that drives inflation, $\rho_{\rm inf} \sim H^2 M_{p}^2$, in order to have a successful inflationary scenario. The integration in Eq.~\eqref{GW-energy} is taken over all the k's in the Hubble patch for which the classical initial configuration is nonzero, i.e. $k_i,~k_f\gg H$. Hence in order to suppress the backreaction of such classical gravity wave configuration on inflation, one has to make sure that $\left|\alpha_{\mathbf k,\lambda}\right|^2+\left|\beta_{\mathbf k,\lambda}\right|^2\ll 1$, or equivalently $\left|\alpha_{\mathbf k,\lambda}\right|\ll 1$ and $\left|\beta_{\mathbf k,\lambda}\right|\ll 1$. In order to see how small these parameters should be, let us assume that $\alpha_{\mathbf k,\lambda}$ and $\beta_{\mathbf k,\lambda}$ are almost scale-independent and the profile of gravity wave spans at least for $\Delta k\simeq H$~\footnote{If these assumptions do not hold, {\it i.e.} the Bogoliubov coefficients are scale-dependent or the observed profile of gravity waves only peaks around some specific value of $k$, then the classical configuration could be easily distinguishable from that of the quantum one.}. Thus, $\langle T_{00} \rangle$ is of order
  \begin{equation}\label{t00}
  \langle T_{00} \rangle \simeq (|\alpha_{\mathbf k,\lambda}|^2 + |\beta_{\mathbf k,\lambda}|^2)H^2 M_{p}^2.
  \end{equation}
  This gravitational energy can at most be of order of the kinetic energy of the inflaton, $\langle T_{00} \rangle \lesssim \epsilon H^2 M_{p}^2$, where $\epsilon \ll 1$ is the slow-roll parameter. Inflation typically ends when $\epsilon \sim {\cal O}(1)$.
   This means that both $|\alpha_{\mathbf k,\lambda}|$ and  $|\alpha_{\mathbf k,\lambda}|$ should be at least smaller than $\sqrt{\epsilon}$ in the slow-roll dominated inflation, which suggests that the amplitude of the classical gravitational waves at large scales should be suppressed by a factor of $\epsilon$ compared to that of Eq.~(\ref{quantum-case}).  This in turn implies that the amplitude of the classical gravity waves in Eq. \eqref{p-classical} is too small to be detected. 

To illustrate our point further, let us consider an alternative possible 
source of classical gravitational waves, namely, the collision of true-vacuum 
bubbles at the end of a first-order cosmological phase 
transition~\cite{turner}. The fraction of vacuum energy released into the low-frequency gravity waves is roughly given by $I(f)\simeq I_{\rm max}(f/f_{\rm max})^3$ for $f\leq f_{\rm max}$, where $f_{\rm max}\simeq 0.2\beta$ is the peak frequency, and $I_{\rm max}\simeq 6\times 10^{-2}\chi$, where $\chi=e^{-N_i}$ and $N_i$ is the number of initial e-foldings before it goes outside the horizon. 
The gravitational energy density of the modes inside the horizon decreases as $a^{-4}$ while outside the horizon, it goes as $a^{-2}$. On the other hand, the frequency redshifts like $a^{-1}$. Thus, the fraction of gravitational energy density today can be written as 
\begin{eqnarray}
\Omega_{\rm GW} h^2 \approx I_{\rm max}\left(e^{-N_i}\right)^{-4}\left(e^{N_f-N_i}\right)^{-2}\left(\frac{T_{\rm RH}}{T_0}\right)^{-4}, 
\label{eq:omega}
\end{eqnarray}
where $N_f\approx 60$ is the total number of e-foldings required 
for a typical high-scale inflationary model, and $T_0=2.3\times 10^{-4}$ eV is today's temperature of the universe. Using the critical value of $\chi\leq \chi_c=1/3$ to achieve percolation and thermalization of the bubbles~\cite{turner}, we find that $N_i\leq 1$. For the conservative lower limit on the reheating temperature $T_{\rm RH}\geq 1$ MeV to have successful nucleosynthesis, we obtain from Eq.~(\ref{eq:omega}): $\Omega h^2\lsim 6\times 10^{-94}$. This is related to the amplitude of the power spectrum for the primordial (tensor and scalar) 
density perturbations by~\cite{kamion}
\begin{eqnarray}
\Omega_{\rm GW} h^2 = A_{\rm GW}P_t=A_{\rm GW}rP_s, 
\label{eq:omega2}
\end{eqnarray}  
where $A_{\rm GW}=2.74\times 10^{-6}(100/g_*)^{1/3}$. For 
$P_s=2.45\times 10^{-9}$~\cite{planck2}, Eqs. (\ref{eq:omega}) and (\ref{eq:omega2}) imply that the tensor-to-scalar ratio, $r\lsim 3\times 10^{-80}$ which is obviously beyond detection. Note that even in the most optimistic scenario 
where all the vacuum energy is tranferred to the gravitational wave which could happen for instance in many preheating models of inflation, we estimate that $r$ is of the same order, and hence, can never be observed. Hence, in classical gravity there will be {\it no} detectable $B$-mode polarization in the CMB spectrum.

In conclusion, in a purely classical theory of gravity, there is {\it no} upper bound on the scale of inflation, and hence, very high-scale
(GUT-scale or even beyond) models of inflation can still be compatible with all the observational constraints. In addition, the amplitude of the primordial
tensor modes of a classical gravity wave is too small to be ever observable in the CMB spectrum. Hence, a positive detection of the $B$-mode polarization in the
CMB spectrum would have profound implications, which will not only
put inflation on firm footing, but will also shed light on the very nature of
the fabric of space-time. 
This will be a huge step forward in resolving the long-standing issue of
whether the space-time gravity should be treated as classical or quantum in nature.

This work is supported by the Lancaster-Manchester-Sheffield Consortium for Fundamental Physics under STFC grant ST/J000418/1. We would like to thank Tirthabir Biswas, Tomi Koivisto and M. M. Sheikh-Jabbari for helpful discussions.

\col{{\bf Note Added:} The BICEP2 collaboration has just announced the detection of the $B$-mode 
polarization at a significance of $> 5\sigma$~\cite{bicep}. According to the 
results of our paper, this is clearly the {\it first direct evidence for the quantum nature of gravity}.  We also note that, after several months our paper first appeared on arXiv, a very similar argument on the quantization of gravity was made by  L.~M.~Krauss and F.~Wilczek,
  Phys.\ Rev.\ D {\bf 89}, 047501 (2014)
  [arXiv:1309.5343 [hep-th]].}

\end{document}